\documentclass[aps,prd,superscriptaddress,showpacs,preprintnumbers]{revtex4}
\usepackage{graphicx,color}
\newcommand{\be}{\begin{equation}}
\newcommand{\ee}{\end{equation}}
\newcommand{\bea}{\begin{eqnarray}}
\newcommand{\eea}{\end{eqnarray}}

\newcommand{\bfk}{\mbox{\boldmath $k$}}

\def\kt{k_\perp}
\newcommand{\bfp}{\mbox{\boldmath $p$}}

\newcommand{\bfP}{\mbox{\boldmath $P$}}

\newcommand{\pup}{p^\uparrow}

\newcommand{\qup}{q^\uparrow}
\newcommand{\qdown}{q^\downarrow}

\newcommand{\la}{\lambda}

\def\lsim{\mathrel{\rlap{\lower4pt\hbox{\hskip1pt$\sim$}}\raise1pt\hbox{$<$}}}
\def\gsim{\mathrel{\rlap{\lower4pt\hbox{\hskip1pt$\sim$}}\raise1pt\hbox{$>$}}}
\def\nostrocostruttino#1\over#2{\mathrel{\mathop{\kern 0pt \rlap
{\hbox{$#1$}}} \hbox{\kern-.135em $#2$}}}

\begin{document}

\title{On the role of Collins effect in the Single Spin Asymmetry $A_N$
in $\pup \, p \to h \, X$ processes}

\author{M.~Anselmino}
\affiliation{Dipartimento di Fisica Teorica, Universit\`a di Torino,
             Via P. Giuria 1, I-10125 Torino, Italy}
\affiliation{INFN, Sezione di Torino, Via P. Giuria 1, I-10125 Torino, Italy}
\author{M.~Boglione}
\affiliation{Dipartimento di Fisica Teorica, Universit\`a di Torino,
             Via P. Giuria 1, I-10125 Torino, Italy}
\affiliation{INFN, Sezione di Torino, Via P. Giuria 1, I-10125 Torino, Italy}
\author{U.~D'Alesio}
\affiliation{Dipartimento di Fisica, Universit\`a di Cagliari, Cittadella
             Universitaria, I-09042 Monserrato (CA), Italy}
\affiliation{INFN, Sezione di Cagliari,
             C.P. 170, I-09042 Monserrato (CA), Italy}
\author{E.~Leader}
\affiliation{Imperial College London - South Kensington Campus,
Prince Consort Road, London SW7 2AZ - U.K.}
\author{S.~Melis}
\affiliation{European Centre for Theoretical Studies in Nuclear Physics
             and Related Areas (ECT*), \\
             Villa Tambosi, Strada delle Tabarelle 286, I-38123 Villazzano,
             Trento, Italy}
\author{F.~Murgia}
\affiliation{INFN, Sezione di Cagliari,
             C.P. 170, I-09042 Monserrato (CA), Italy}
\author{A.~Prokudin}
\affiliation{Jefferson Laboratory, 12000 Jefferson Avenue, Newport News,
VA 23606, USA}
\date{\today}

\begin{abstract}

The much debated issue of the transverse single spin asymmetry $A_N$
observed in the inclusive large $P_T$ production of a single hadron in $p\,p$
interactions, $\pup \,p \to \pi \, X$, is considered in a TMD factorization
scheme. A previous result \cite{Anselmino:2004ky,Anselmino:2005sh} stating
that the maximum contribution of the Collins effect is strongly suppressed,
is revisited, correcting a numerical error. New estimates are given,
adopting the Collins functions recently extracted from SIDIS and $e^+e^-$ data,
and phenomenological consequences are discussed.

\end{abstract}


\maketitle

\section{\label{1}Introduction and formalism}

The understanding of spin effects, and in particular, for parity conserving
processes, of transverse Single Spin Asymmetries (SSAs) has always been one of
the major challenges for QCD and any fundamental quantum field theory. Such
effects, abundantly observed in experiments, are not generated by the
perturbative dynamical properties of the Standard Model elementary
interactions and originate from more profound phenomena related to the
intrinsic structure of the nucleons. Thus, the study of these often neglected
effects has recently opened a new phase in our exploration of the partonic
composition of hadrons.

Most progress has occurred in the study of the rich azimuthal dependences
measured in Semi-Inclusive Deep Inelastic Scattering (SIDIS) of leptons off
polarized nucleons by dedicated experiments, HERMES (DESY), COMPASS (CERN)
and JLab. These SIDIS azimuthal asymmetries are interpreted and discussed
in terms of new, unintegrated, Transverse Momentum Dependent distribution
and fragmentation functions (shortly, TMDs); these offer new information on
the properties of quarks and gluons, which go well beyond the usual
one-dimensional description of Partonic Distribution Functions (PDFs) in
terms of longitudinal momentum fraction only.

In particular the Sivers distributions \cite{Sivers:1989cc, Sivers:1990fh,
Boer:1997nt} and the Collins fragmentation functions \cite{Collins:1992kk}
have been extracted \cite{Anselmino:2005ea,Anselmino:2008sga, Vogelsang:2005cs,
Collins:2005ie,Anselmino:2005an,Efremov:2006qm} from SIDIS data, and, thanks
to complementary information from Belle on the Collins function
\cite{Abe:2005zx,Seidl:2008xc}, a first extraction of the transversity
distribution has been possible \cite{Anselmino:2007fs,Anselmino:2008jk}.

All these analyses have been performed in the $\gamma^* - p$ center of mass
(c.m.) frame, within a QCD factorization scheme, according to which the
SIDIS cross section is written as a convolution of TMDs and elementary
interactions:
\be
d\sigma^{\ell p \to \ell' h X} = \sum_q \hat f_{q/p}(x, \bfk_\perp; Q^2)
\otimes d\hat\sigma^{\ell q \to \ell q} \otimes
\hat D_{h/q}(z, \bfp_\perp; Q^2) \>,
\ee
where $\bfk_\perp$ and $\bfp_\perp$ are, respectively, the transverse
momentum of the quark in the proton and of the final hadron with respect
to the fragmenting quark. At order $k_\perp/Q$ the observed $P_T$ of the
hadron is given by
\be
\bfP_T = z \, \bfk_\perp + \bfp_\perp  \>.
\ee
There is a general consensus \cite{Ji:2004wu,Ji:2004xq,Bacchetta:2008xw}
that such a scheme holds in the kinematical region defined by
\be
P_T \simeq k_\perp \simeq \Lambda_{\rm QCD} \ll Q \>.
\ee
The presence of the two scales, small $P_T$ and large $Q$, allows to
identify the contribution from the unintegrated partonic distribution
($P_T \simeq \kt$), while remaining in the region of validity of the QCD
parton model. At larger values of $P_T$ other mechanisms, like quark-gluon
correlations and higher order perturbative QCD contributions become important
\cite{Ji:2006br,Anselmino:2006rv,Bacchetta:2008xw}. A similar situation
\cite{Collins:1984kg,Ji:2004xq,Ji:2006ub,Ji:2006vf,Arnold:2008kf,
Anselmino:2002pd,Anselmino:2009st}
holds for Drell-Yan processes, $A B \to \ell^+ \ell^- X$, where the two scales
are the small transverse momentum, $q_T$, and the large invariant mass, $M$,
of the dilepton pair.

The situation is not so clear for processes in which only one large scale is
detected, like the inclusive production, at large $P_T$, of a single particle
in hadronic interactions, $A B \to C X$. However, the most striking and large
SSAs have been \cite{Adams:1991rw,Adams:1991cs,Adams:1991ru,Bravar:1996ki}
and keep being measured \cite{Adams:2003fx,Adler:2005in,Abelev:2008af,
Lee:2007zzh, Arsene:2008aa, Aidala:2008qj} in these reactions. The TMD
factorization for these processes
was first suggested in Refs.~\cite{Sivers:1989cc,Sivers:1990fh} and adopted
in Refs.~\cite{Anselmino:1994tv,Anselmino:1998yz,Anselmino:1999pw}
to explain the large single spin asymmetries observed by the E704
Collaboration \cite{Adams:1991cs}. The same approach led to
successful predictions \cite{D'Alesio:2004up,D'Alesio:2007jt} for the values
of $A_N$ measured at RHIC \cite{Nogach:2006gm}.

Alternative approaches to explain the origin of SSAs, linking collinear
partonic dynamics to higher-twist quark-gluon correlations, were originally
proposed in Refs.~\cite{Efremov:1981sh,Efremov:1984ip,Qiu:1991pp,Qiu:1998ia,Kanazawa:2000hz}
and phenomenologically used in Refs.
\cite{Kouvaris:2006zy,Eguchi:2006qz,Eguchi:2006mc,Koike:2009ge}.
These two approaches, the TMD factorization and the higher-twist correlations,
have been shown to be related \cite{Boer:2003cm,Yuan:2009dw}
and consistent with each other \cite{Ji:2006ub,Ji:2006vf,Koike:2007dg}.

However, a definite proof of the validity of the TMD factorization for
hadronic inclusive processes with one large scale only is still lacking.
Due to this, the study of dijet production at large $P_T$ in hadronic
processes was proposed \cite{Boer:2003tx,Bomhof:2004aw,Bacchetta:2005rm,
Bomhof:2007su}, where the second small scale is the total $q_T$
of the two jets, which is of the order of the intrinsic partonic momentum
$k_\perp$. This approach leads to a modified TMD factorization approach,
with the inclusion in the elementary processes of gauge link color factors
\cite{Bomhof:2006dp,Bacchetta:2007sz,Ratcliffe:2007ye}. Despite the
identification of two separate scales, some problems with the TMD
factorization for hadronic processes with the final inclusive production
of two jets, or two hadrons, have been recently pointed
out~\cite{Rogers:2010dm,Mulders:2011zt}. TMD factorization is expected to
work for the observation, inside a jet with large transverse momentum $P_T$,
of a final hadron with a transverse momentum with respect to the jet
direction, like $\pup p \to {\rm jet} + \pi + X$ 
\cite{Yuan:2007nd,Yuan:2008yv,D'Alesio:2010am}.

In this paper we consider SSAs in $\pup \, p \to \pi \, X$ processes, with
only one large $P_T$ final pion detected, for which data are available.
We adopt the TMD factorization scheme \cite{Anselmino:2004ky,
Anselmino:2005sh, Anselmino:1994tv,Anselmino:1998yz,Anselmino:1999pw,
D'Alesio:2004up,D'Alesio:2007jt},
\be
d\sigma^{p p \to \pi X} = \sum_{a,b,c,d} \hat f_{a/p}(x_a, \bfk_{\perp a};
Q^2) \otimes \hat f_{b/p}(x_b, \bfk_{\perp b}; Q^2) \otimes
d\hat\sigma^{ab \to cd} \otimes \hat D_{\pi/c}(z, \bfp_\perp; Q^2) \>,
\label{fact}
\ee
as a natural phenomenological extension of the corresponding collinear
factorization, based on the convolution of integrated parton
distributions (PDFs) and fragmentation functions (FFs) with QCD elementary
dynamics; this collinear factorization works well in computing unpolarized
cross sections, but fails in explaining SSAs as there is no single spin
effect in the collinear PDFs and FFs, and in lowest order QCD dynamics.
In the polarized case, for a generic process $(A,S_A) + (B,S_B) \to C + X$,
Eq.~(\ref{fact}) explicitely reads~\cite{Anselmino:2005sh}:
\bea
\frac{E_C \, d\sigma^{(A,S_A) + (B,S_B) \to C + X}}
{d^{3} \bfp_C} = \!\!\!\!\! \sum_{a,b,c,d, \{\la\}}
&& \!\!\! \int \frac{dx_a \,
dx_b \, dz}{16 \pi^2 x_a x_b z^2  s} \;
d^2 \bfk_{\perp a} \, d^2 \bfk_{\perp b}\, d^3 \bfp_{\perp}\,
\delta(\bfp_{\perp} \cdot \hat{\bfp}_c) \, J(p_\perp)
\nonumber \\
&\times& \rho_{\la^{\,}_a,
\la^{\prime}_a}^{a/A,S_A} \, \hat f_{a/A,S_A}(x_a,\bfk_{\perp a})
\> \rho_{\la^{\,}_b, \la^{\prime}_b}^{b/B,S_B} \,
\hat f_{b/B,S_B}(x_b,\bfk_{\perp b}) \label{gen1} \\
&\times& \hat M_{\la^{\,}_c, \la^{\,}_d; \la^{\,}_a, \la^{\,}_b} \,
\hat M^*_{\la^{\prime}_c, \la^{\,}_d; \la^{\prime}_a,
\la^{\prime}_b} \> \delta(\hat s + \hat t + \hat u) \> \hat
D^{\la^{\,}_C,\la^{\,}_C}_{\la^{\,}_c,\la^{\prime}_c}(z,\bfp_\perp)
\>. \nonumber
\eea
Further details and a full explanation of the notations can be found in
Refs.~\cite{Anselmino:2004ky,Anselmino:2005sh} (where $\bfp_\perp$ is
denoted as $\bfk_{\perp C}$).

There are many contributions, from different TMDs, to the SSAs according
to the above expression. In Ref.~\cite{Anselmino:2004ky} it was argued
that only the Sivers effect contributes significantly; a further small
contribution from the Collins effect is possible (largely suppressed by
phase integrations), while all other TMD contributions are utterly
negligible. However that conclusion was affected by a wrong sign
\cite{Yuan:2008tv, Anselmino:2008uy} in one of the elementary interactions
and the Collins effect contribution was underestimated. It remains true
that all other contributions to the SSAs are negligible.

\vskip 12pt
\centerline{\bf Reconsideration of the Collins contribution}
\vskip 12pt

We reconsider here the Collins contribution to the SSA
\be
A_N = \frac{d\sigma^\uparrow - d\sigma^\downarrow}
           {d\sigma^\uparrow + d\sigma^\downarrow} \label{an}
%
\quad\quad {\rm where} \quad\quad
%
d\sigma^{\uparrow, \downarrow} \equiv
\frac{E_\pi \, d\sigma^{p^{\uparrow, \downarrow} \, p \to \pi \, X}}
{d^{3} \bfp_\pi} \>\cdot
\ee
Such a contribution can be computed according to the TMD factorized
expression \cite{Anselmino:2004ky,Anselmino:2005sh}:
\bea
[d\sigma^\uparrow - d\sigma^\downarrow]_{\rm Collins}
&=& \!\!\! \sum_{q_a,b,q_c,d} \int \frac{dx_a \, dx_b \, dz}
{16 \, \pi^2 \, x_a \, x_b \, z^2 s} \; d^2 \bfk_{\perp a} \,
d^2 \bfk_{\perp b}\, d^3 \bfp_{\perp}\,
\delta(\bfp_\perp \cdot \hat{\bfp}_c) \> J(p_{\perp}) \>
\delta(\hat s + \hat t + \hat u) \nonumber \\
&\times& \Delta_Tq_a(x_a, k_{\perp a}) \,
\cos (\phi_a + \varphi_1 - \varphi_ 2 + \phi_\pi^H) \label{numanc} \\
&\times& f_{b/p}(x_b, k_{\perp b}) \>
\left[ \hat M_1^0 \, \hat M_2^0 \right]_{q_ab\to q_cd} \>
\Delta^N D_{\pi/\qup_c}(z, p_\perp) \>, \nonumber
\eea
which can be easily interpreted: the transversity distribution
$\Delta_Tq_a$ (or $h_1^{q_a}$) of the quark $q_a$, couples with the
unpolarized TMD for parton $b$, $f_{b/p}$, and the Collins fragmentation
function $\Delta^N D_{\pi/\qup_c} = (2p_\perp/zM_h) \,
H_{1}^{\perp q_c}$~\cite{Bacchetta:2004jz} of quark $q_c$ into a $\pi$.
The $\hat M_i^0$ are the independent helicity amplitudes defined in Refs.~\cite{Anselmino:2004ky,Anselmino:2005sh},
describing the lowest order QCD interactions and the quantity
$\hat M_1^0 \, \hat M_2^0$ is proportional to the elementary spin transfer
cross-section $d\hat\sigma^{\qup_a b \to \qup_c d} - d\hat\sigma^{\qup_a b \to \qdown_c d}$. The (suppressing) phase factor $\cos(\phi_a + \varphi_1 -
\varphi_ 2 + \phi_\pi^H)$ originates from the $\bfk_\perp$ dependence of
the unintegrated transversity distribution, the polarized elementary
interaction and the spin-$\bfp_\perp$ correlation in the Collins function.
The explicit expressions of $\varphi_1, \varphi_2$ and $\phi_\pi^H$ in terms
of the integration variables can be found via Eqs.~(60)-(63) in
\cite{Anselmino:2005sh} and Eqs.~(35)-(42) in  \cite{Anselmino:2004ky}.

The denominator of Eq.~(\ref{an}) is twice the unpolarized cross-section
and is given in our TMD factorization scheme by:
\bea
[d\sigma^\uparrow + d\sigma^\downarrow]
&=& 2 \, d\sigma^{\rm unp} = 2 \, \frac{E_\pi \, d\sigma^{p \, p \to \pi \, X}}
{d^{3} \bfp_\pi} \nonumber \\
&=& \!\!\! \sum_{a,b,c,d} \int \frac{dx_a \, dx_b \, dz}
{16 \, \pi^2 \, x_a \, x_b \, z^2 s} \; d^2 \bfk_{\perp a} \,
d^2 \bfk_{\perp b}\, d^3 \bfp_{\perp}\,
\delta(\bfp_\perp \cdot \hat{\bfp}_c) \> J(p_{\perp}) \>
\delta(\hat s + \hat t + \hat u) \label{den} \\
&\times& f_{a/p}(x_a, k_{\perp a}) \> f_{b/p}(x_b, k_{\perp b}) \,
\left[ |\hat M_1^0|^2 + |\hat M_2^0|^2 + |\hat M_3^0|^2 \right]_{ab\to cd}
\> D_{\pi/c}(z, p_\perp) \>. \nonumber
\eea

The explicit expressions of $\hat M_1^0 \, \hat M_2^0$, which give the
QCD dynamics in Eq.~(\ref{numanc}), can be found, for all possible
elementary interactions, in Ref.~\cite{Anselmino:2005sh}. Unfortunately,
it turns out that, for the important $q g \to q g$ channel, as given in
the last line of Eq.~(55) in Ref.~\cite{Anselmino:2004ky} and in the last
term of Eq.~(71) in Ref.~\cite{Anselmino:2005sh}, there is an overall
wrong sign. Moreover, at the time of the numerical estimates of
Ref.~\cite{Anselmino:2004ky}, no information was available on the
transversity distribution and the Collins functions, and we computed
the maximum Collins contribution to $A_N$ by replacing these unknown
functions with their upper limits imposed by positivity conditions.
Much more information is now available on the transversity distribution
and the Collins fragmentation function from studies of SIDIS data
by COMPASS and HERMES Collaborations and $e^{+} e^{-}$ data by Belle
Collaboration \cite{Anselmino:2007fs,Anselmino:2008jk}. Notice that the
Collins fragmentation function is expected to be process independent
\cite{Collins:2004nx,Yuan:2008yv,Yuan:2009dw}, so that this new information
can be used in the inclusive hadronic reactions we consider here.

Due to these reasons -- the correction of a previous numerical error,
and the use of actual information on the functions contributing to
Eq.~(\ref{numanc}) -- we are now able to attempt new realistic estimates
of the contribution of the Collins effect to the intriguing data on $A_N$.

\section{The role of the Collins effect\label{sec:scan}}

Thus, in this section, we investigate in much detail the phenomenology of
the Collins effect for pion SSAs at STAR, BRAHMS and E704 kinematics,
according to Eqs.~(\ref{an})--(\ref{den}), exploiting all the available and
updated information on the transversity  distribution and the Collins
fragmentation functions coming from SIDIS and $e^+e^-$ annihilation data.
In addition, we correct the numerical error in one of the
elementary interactions mentioned above, which affected the conclusions
of Ref.~\cite{Anselmino:2004ky}.

The first combined extraction of the quark transversity distribution and
the Collins function was presented in Ref.~\cite{Anselmino:2007fs}. We will
refer to it as the ``SIDIS-1" fit. In this extraction, the Kretzer set of
unpolarized FFs~\cite{Kretzer:2000yf} was adopted. An updated extraction
of the transversity and Collins functions was presented in
Ref.~\cite{Anselmino:2008jk}. We will refer to the corresponding set of
parameterizations as the ``SIDIS-2" fit. In this case, the set of pion and
kaon FFs by de Florian, Sassot and Stratmann~\cite{deFlorian:2007aj}, which
became available at that time, was considered.

Let us recall the main features of the parameterizations adopted in
Refs.~\cite{Anselmino:2007fs,Anselmino:2008jk}. The analysis of SIDIS and
$e^+e^-$ data is performed at leading order, ${\cal O}(k_\perp/Q)$, in the
TMD factorization approach, where $Q$ is the large scale in the process.
A simple factorized form of the TMD functions was adopted, using a Gaussian
shape for their $k_\perp$ dependent component. For the unpolarized parton
distribution and fragmentation functions we have:
\begin{equation}
f_{q/p}(x,k_\perp) = f_{q/p}(x)\,
\frac{e^{-k_\perp^2/\langle k_\perp^2 \rangle}}
{\pi \langle k_\perp^2 \rangle}\,,
\label{eq:pdf-unp}
\quad\quad\quad
D_{h/q}(z,p_\perp) = D_{h/q}(z)\,
\frac{e^{-p_\perp^2/\langle p_\perp^2 \rangle}}
{\pi \langle p_\perp^2 \rangle}\,,
\label{eq:ff-unp}
\end{equation}
where $\langle k_\perp^2\rangle$ and  $\langle p_\perp^2 \rangle$
have been fixed by analyzing the Cahn effect in unpolarized SIDIS processes,
see Ref.~\cite{Anselmino:2005ea}:
\begin{equation}
\langle k_\perp^2\rangle = 0.25\, {\rm GeV}^2\,, \qquad\qquad
\langle p_\perp^2\rangle = 0.20\, {\rm GeV}^2\,.
\label{eq:k-p-cahn}
\end{equation}

For the usual integrated PDFs $f_{q/p}(x)$ we adopted the GRV98
set~\cite{Gluck:1998xa} and, as said above, for the integrated FFs
$D_{h/q}(z)$ we used the Kretzer set~\cite{Kretzer:2000yf} for the
SIDIS-1 fit and the DSS one~\cite{deFlorian:2007aj} for the SIDIS-2 fit.
We have taken into account their DGLAP QCD evolution.

The quark transversity distribution, $\Delta_T q(x,k_\perp)$, and the
Collins fragmentation function, $\Delta^N D_{h/q^\uparrow}(z,p_\perp)$,
have been parametrized as follows:
\begin{equation}
\Delta_T q(x,k_\perp) = \frac{1}{2}\,{\cal N}_q^T(x)\,\left[\,f_{q/p}(x)+
\Delta q(x)\,\right]\,\frac{e^{-k_\perp^2/\langle k_\perp^2 \rangle_T}}
{\pi \langle k_\perp^2 \rangle_T}\,,
\label{eq:transv-par}
\end{equation}
\begin{equation}
\Delta^N\! D_{h/q^\uparrow}(z,p_\perp) = 2 {\cal N}_q^C(z)\,D_{h/q}(z)\,
h(p_\perp)\,\frac{e^{-p_\perp^2/\langle p_\perp^2 \rangle}}
{\pi \langle p_\perp^2 \rangle}\,,
\label{eq:coll-par}
\end{equation}
where $\Delta q(x)$ is the usual collinear quark helicity distribution,
\begin{equation}
{\cal N}_q^T(x) = N_q^T x^{\alpha_q}(1-x)^{\beta_q}\,
\frac{(\alpha_q+\beta_q)^{(\alpha_q+\beta_q)}}
{\alpha_q^{\alpha_q}\beta_q^{\beta_q}}\,,
\label{eq:nq-trans}
\end{equation}
and
\begin{equation}
{\cal N}_q^C(z) = N_q^C z^{\gamma_q}(1-z)^{\delta_q}\,
\frac{(\gamma_q+\delta_q)^{(\gamma_q+\delta_q)}}
{\gamma_q^{\gamma_q}\delta_q^{\delta_q}}\,,
\label{eq:nq-coll}
\end{equation}
with $|N_q^{T(C)}|\leq 1$. Moreover,
\begin{equation}
h(p_\perp) = \sqrt{2e}\,\frac{p_\perp}{M_h}\,e^{-p_\perp^2/M_h^2}\,.
\label{eq:h-coll}
\end{equation}

With these choices, the transversity and Collins functions automatically
fulfill their proper Soffer and positivity bounds respectively,
for any values of the $(x,k_\perp)$ and $(z,p_\perp)$ variables.
The quark helicity distributions $\Delta q(x)$, required for the Soffer
bound, are taken from Ref.~\cite{Gluck:2000dy}. The term
$[f_{q/p}(x) +\Delta q(x)]$ in Eq.~(\ref{eq:transv-par}) is evaluated at the initial
scale and evolved at the appropriate $Q^2$ values using the
transversity evolution kernel. Similarly, for the $Q^2$ evolution of the
Collins function, which remains so far unknown, we considered the
unpolarized DGLAP evolution of its collinear factor $D_{h/q}(z)$.

Despite the simplicity of these functional forms, they still involve,
in the most general case, a huge number of free parameters.
In Refs.~\cite{Anselmino:2007fs,Anselmino:2008jk} we therefore
adopted some additional, physically motivated assumptions,
in order to keep the number of free parameters reasonably low.
First of all, for the transversity distribution we used only valence
quark contributions. In addition, for the fragmentation functions we
considered two different expressions for ${\cal N}_q^C$, corresponding
to the so-called ``favoured" and ``unfavoured" FFs,
${\cal N}_{\rm fav}^C(z)$ and ${\cal N}_{\rm unf}^C(z)$; for example,
for pions, we had:
\begin{equation}
{\cal N}^C_{\pi^+/u}(z) = {\cal N}^C_{\pi^+/\bar{d}}(z) =
 {\cal N}^C_{\pi^-/\bar{u}}(z) =
 {\cal N}^C_{\pi^-/d}(z) = {\cal N}^C_{\rm fav}(z)\,,
\label{eq:d-pi-fav}
\end{equation}
\begin{equation}
{\cal N}^C_{\pi^+/\bar{u}}(z) = {\cal N}^C_{\pi^+/d}(z) =
{\cal N}^C_{\pi^-/u}(z) =
{\cal N}^C_{\pi^-/\bar{d}}(z) = {\cal N}^C_{\pi^\pm/s}(z) =
{\cal N}^C_{\pi^\pm/\bar{s}}(z) = {\cal N}^C_{\rm unf}(z)\,.
\label{eq:d-pi-unf}
\end{equation}

Notice, however, that our complete parameterization of the Collins FFs, Eq.~(\ref{eq:coll-par}), allows for further differences among parton
flavours, possibly contained in the usual unpolarized FFs.

In addition, we kept a flavour dependence in the coefficients $N_{u,d}^T$
and $N_{\rm fav, unf}^C$, while the parameters $\alpha_q$, $\beta_q$,
$\gamma_q$, $\delta_q$ and $M_h$ were taken to be flavour independent.
For simplicity we also assumed that
$\langle k_\perp^2\rangle_T = \langle k_\perp^2\rangle$.
With these choices, we were left with a total of 9 free parameters for the
SIDIS-1 and SIDIS-2 fit parameterizations:
\begin{equation}
N^T_{u},\, N^T_{d},\, N^C_{\rm fav},\, N^C_{\rm unf},\,
\alpha,\, \beta,\, \gamma,\, \delta, M_h \,.
\label{eq:9-par}
\end{equation}

Both fits gave good results. However, a study of the statistical
uncertainties of the best fit parameters, and a comparison of the two sets
of parameterizations, SIDIS-1 and SIDIS-2, clearly shows that SIDIS data
are not presently able to constrain the large $x$ behaviour of the quark
($u$, $d$) transversity distributions, leaving a large uncertainty in the
possible values of the parameter $\beta$. In fact, the range of Bjorken $x$
values currently explored by HERMES and COMPASS experiments is limited to
$x_B \lesssim 0.3$.

This uncertainty in the knowledge of the transversity distribution at
large $x$ values has relevant consequences when one uses the
parameterizations extracted from SIDIS and $e^+e^-$ data for the study
of single spin asymmetries in hadronic collisions. In this case the largest
pion asymmetries have been measured at large Feynman $x$ values,
$x_F \gsim 0.3$; then, kinematical cuts imply that the
transversity distribution is probed at even larger $x$ values.

In order to assess the possible relevance of the Collins effect in explaining
a large value of $A_N$ in $pp$ collisions we should explore in greater
details the large $x$ contribution of the transversity distribution.
We have then devised a simple analysis, to which we will refer to as the
``scan procedure" and which is based on the following considerations.

\begin{itemize}
\item
In our parameterizations, the large $x$ behaviour of the quark tranversity
distributions is driven by the parameters $\beta_q$, the exponents
of the $(1-x)$ factor in ${\cal N}_q^T(x)$, see Eqs.~(\ref{eq:transv-par})
and (\ref{eq:nq-trans}). Not surprinsingly, for the dominant $u$ and $d$
contributions, the values of the single $\beta_u = \beta_d = \beta$
parameter are indeed very different in the SIDIS-1 and SIDIS-2 sets,
despite the fact that they offer comparably good fits of SIDIS and $e^+e^-$
data. As a consequence, the two sets give strongly different
estimates of the pion SSAs at large $x_F$ in hadronic collisions.
It is then natural to conclude that the choice of a flavour-independent
$\beta$ parameter, good for SIDIS data, is a much too strong assumption for
the hadronic collisions, and must be released in this analysis.
We also notice that both in SIDIS-1 and SIDIS-2 fits the best-fit
values of $\delta_{\rm fav} = \delta_{\rm unf} = \delta$ are very close
to zero.

\item
We therefore start the scan procedure by performing a preliminary
9-parameter ``reference fit" to SIDIS and $e^+e^-$ data taking,
w.r.t.~Eq.~(\ref{eq:9-par}), $\beta_u \neq \beta_d$ and $\delta=0$.
We then let the two parameters $\beta_u$ and $\beta_d$ vary independently
over the range $0.0$---$4.0$ by discrete steps of $0.5$. Larger values of
$\beta$ would give negligible contribution to $A_N$. For each of the 81
points in this two-dimensional grid in ($\beta_u$, $\beta_d$) space
we perform a new 7-parameter fit to SIDIS and $e^+e^-$ data, keeping
$\beta_{u,d}$ fixed and $\delta=0$, but leaving all remaining 7 parameters in
Eq.~(\ref{eq:9-par}) free.

\item
As a next step we select only those sets of parameters from the scan
procedure over the $(\beta_u,\beta_d)$ grid leading to an increment of the
total $\chi^2$ of the fit, as compared to the corresponding 9-parameter
reference fit, smaller than a given chosen value, $\Delta\chi^2$. Notice
that, since the reference fit and the scan fits have a different number of
free parameters, the selection criterium is applied to the total $\chi^2$
rather than to the $\chi^2$ per degree of freedom, $\chi^2_{\rm dof}$.
The chosen value of $\Delta\chi^2$ is the same as that used in
Refs.~\cite{Anselmino:2007fs,Anselmino:2008jk} to generate the error band,
following the procedure described in Appendix A of
Ref.~\cite{Anselmino:2008sga}. It is worth noticing that all 81 points
of our grid in $(\beta_u,\beta_d)$ lead to acceptable fits; this further
confirms the observation that the SIDIS data cannot constrain the large $x$
behaviour of the transversity distribution.

\item
For each of the selected sets, we calculate the Collins pion SSA for
polarized $pp$ collisions, Eqs.~(\ref{an})--(\ref{den}), in the kinematical
regions of the available data from the E704 Collaboration at Fermilab and
the STAR (for $\pi^0$) and BRAHMS (for charged pions) Collaborations at RHIC.

Finally we generate a ``scan band", by taking the envelope of ALL curves
for $A_N(\pi)$ obtained by using the sets selected in the scan
procedure, and compare this band with the experimental data available.
This band shows the potentiality of the Collins effect alone to account
for $A_N(p^\uparrow p\to\pi X)$ data while preserving a combined fair
description (quantified by $\Delta\chi^2)$ of the SIDIS and $e^+e^-$ data on
Collins azimuthal asymmetries.

\item
Notice that in the data sets used for the fits of the scan procedure we have
also included the recent preliminary data by the COMPASS Collaboration
on SIDIS off a transversely polarized proton target~\cite{Adolph:2012sn},
which were not available at the time when
Refs.~\cite{Anselmino:2007fs,Anselmino:2008jk} were published. Strictly
speaking, therefore, the parameterizations used
as starting point of the scan procedure are not the same as those published
in Refs.~\cite{Anselmino:2007fs,Anselmino:2008jk}. However, we have verified
that the new parameterizations are only slightly different from the previous
ones and are qualitatively consistent with them. Therefore, here we will
not present and discuss them further, referring for more details to a
future complete upgrade of our parameterizations. Apart from the insertion
of this new set of COMPASS data, all technical aspects of the fitting
procedure followed here are the same as in
Refs.~\cite{Anselmino:2007fs,Anselmino:2008jk}, with the difference that,
for the QCD evolution of the Collins function, limited to
its collinear $z$-dependent $D_{h/q}(z)$ factor, we have attempted, in
addition to an unpolarized-like evolution, also a transversity-like one,
driven by the transversity evolution kernel.
\end{itemize}

\begin{figure}
\begin{minipage}[c]{19cm}
\includegraphics[width=14cm,angle=0]{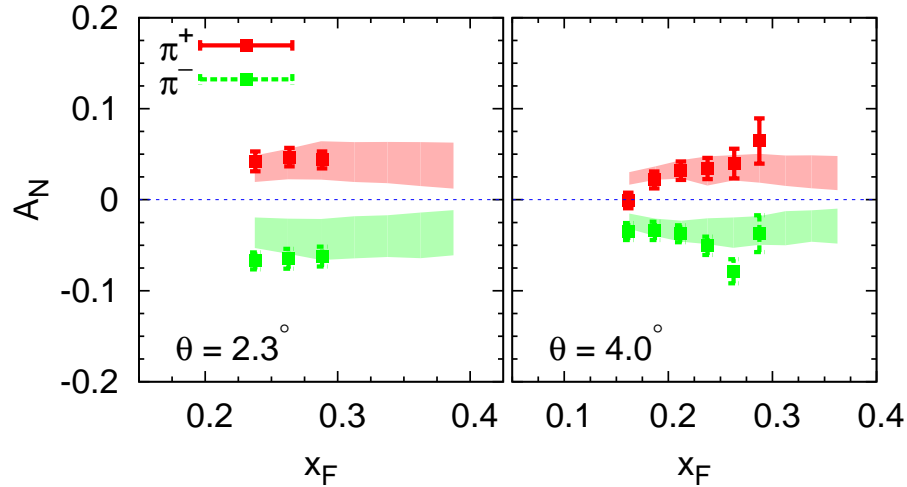}
\end{minipage}
\caption{
Scan band ({\it i.e.} the envelope of possible values) for the Collins
contribution to the charged pion single spin asymmetries $A_N$, as a
function of $x_F$ at two different scattering angles, compared with the
corresponding BRAHMS experimental data \cite{Lee:2007zzh}.
The shaded band is generated, adopting the GRV98 and GRSV2000 sets of
collinear PDFs, the Kretzer FF set and an ``unpolarized-like" evolution
for the Collins function, following the procedure explained in the text.
}
\label{fig:an-brahms}
\end{figure}
\begin{figure}
\begin{minipage}[c]{19cm}
\includegraphics[width=14cm,angle=0]{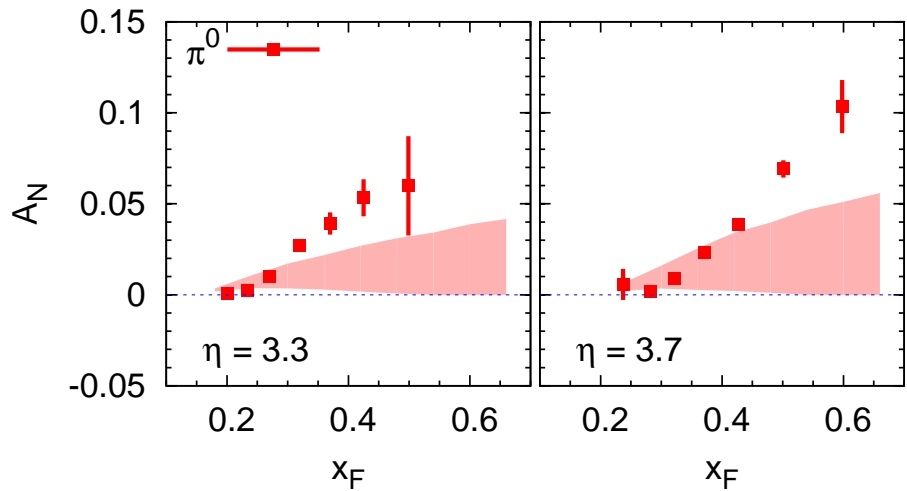}
\end{minipage}
\caption{
Scan band ({\it i.e.} the envelope of possible values) for the Collins
contribution to the neutral pion single spin asymmetry $A_N$, as a
function of $x_F$ at two different rapidity values, compared with the
corresponding STAR experimental data \cite{Abelev:2008af}.
The shaded band is generated, adopting the GRV98 and GRSV2000 sets of
collinear PDFs, the Kretzer FF set and an ``unpolarized-like" evolution
for the Collins function, following the procedure explained in the text.
}
\label{fig:an-star}
\end{figure}

\vskip 12pt
\centerline{\bf Results and comments}
\vskip 12pt

We have computed $A_N (\pup p \to \pi X)$ adopting, as explained above, a
single set of collinear parton distributions~\cite{Gluck:1998xa,Gluck:2000dy},
two different sets for the pion collinear
FFs~\cite{Kretzer:2000yf,deFlorian:2007aj} and two different (partial)
evolution schemes for the Collins function. In Figs. 1-3 we show some of
our results, avoiding the explicit presentation of other cases with
very similar outcomes (some further comments are given below). For all
results presented the Kretzer set for the unpolarized FFs and the
unpolarized-like Collins evolution have been used.

\begin{figure}
\includegraphics[width=14cm,angle=0]{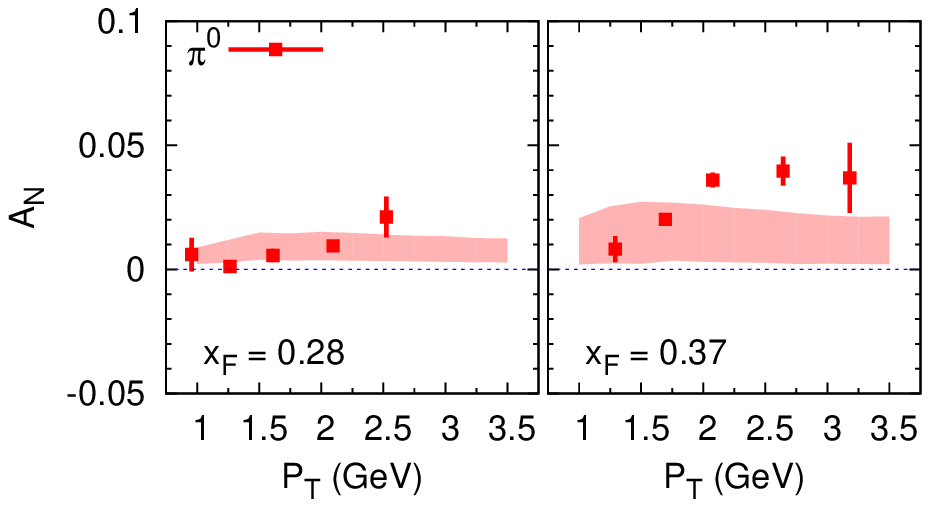}
\\
\includegraphics[width=14cm,angle=0]{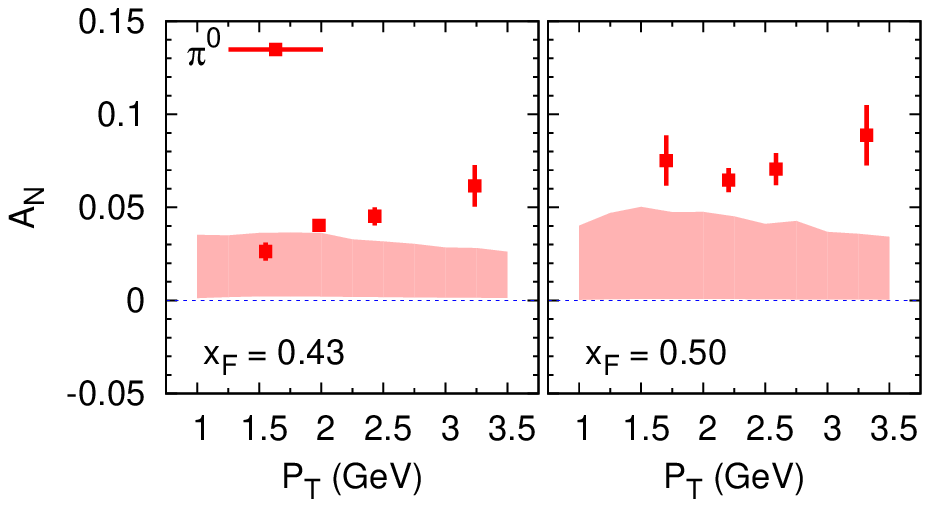}
\caption{
The same as for Fig.~\ref{fig:an-star}, but with the STAR data plotted
vs.~the  pion transverse momentum, $P_T$, for different bins in $x_F$,
$x_F =$ 0.28, 0.37, 0.43 and 0.50. }
\label{fig:an-star-xf}
\end{figure}

In Fig.~\ref{fig:an-brahms} the scan band for $A_N$, as a function
of $x_F$ at fixed scattering angles, is shown for charged pions and
BRAHMS kinematics, while in Fig.~\ref{fig:an-star} the same result is given,
at fixed rapidity values, for neutral pions and STAR kinematics; analogous
results, as a function of $P_T$ at several fixed $x_F$ values,
are shown for STAR kinematics in Fig.~\ref{fig:an-star-xf}.

%

These results allow to draw some first qualitative conclusions:

\begin{itemize}
\item
The Collins contribution to $A_N$ is not as tiny as claimed in
Ref.~\cite{Anselmino:2004ky};
\item
The Collins effect alone might in principle be able to explain the BRAHMS
charged pion results on $A_N$ in the full kinematical range so far explored;
\item
The full amount of the $\pi^0$ STAR data on $A_N$ cannot be explained by the
Collins contribution alone. The Collins effect might be sufficient for the
small $x_F$ portion of the data; however, it is not sufficient for the
medium-large $x_F$ range of STAR data, $x_F \gsim 0.3$.
\end{itemize}

The results obtained with a different choice of the fragmentation functions
(the DSS set) are qualitatively very similar in the large $x_F$ regions.
They are instead smaller in size at smaller $x_F$, due to the
large gluon contribution in the leading order (LO) DSS fragmentation functions. The use of a
transversity-like Collins evolution, rather than the unpolarized one, does
not lead to any significant difference, in all cases.

At this point, in order to fully assess the role of the Collins effect in
understanding the large SSAs for neutral pions measured at large $x_F$ by
the STAR Collaboration at RHIC, we have performed several further tests.

First of all, we should make it clear that the scan bands presented in our
plots have nothing to do with the statistical error bands presented in
Refs.~\cite{Anselmino:2007fs,Anselmino:2008jk}. There, the error bands
are generated by estimating the uncertainty in the best fit values of the
parameters, according to the procedure described in detail
in Appendix A of Ref.~\cite{Anselmino:2008sga}. Instead, the scan bands
in this paper are obtained by simply taking the envelope of all curves
generated by the selected best fit sets within the full grid in $\beta_{u,d}$.

\begin{figure}
\begin{minipage}[c]{19cm}
\includegraphics[width=14cm,angle=0]{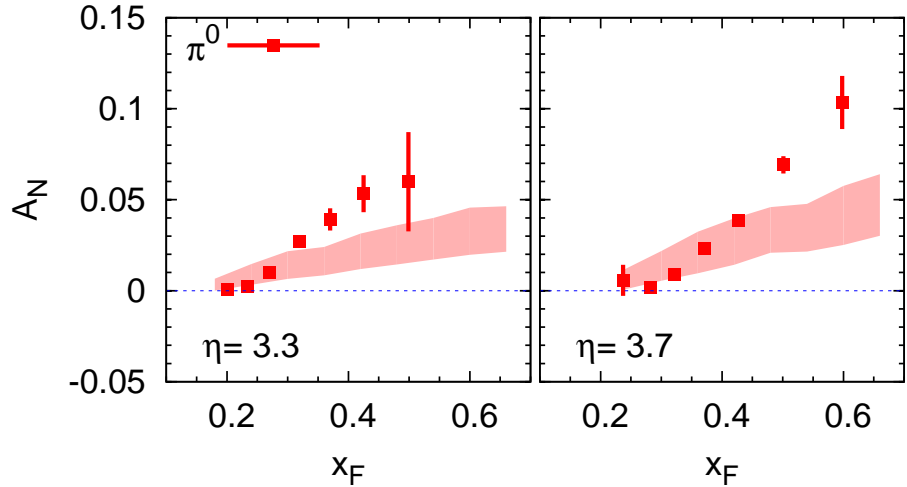}
\end{minipage}
\caption{
The Collins contribution to the neutral pion single spin asymmetry $A_N$,
compared with the corresponding STAR experimental data at two fixed pion
rapidities \cite{Abelev:2008af}. The shaded statistical error band is
generated, adopting the
GRV98 and GRSV2000 sets of collinear PDFs, the Kretzer FF set and an
``unpolarized-like" evolution for the Collins function, starting from
the 7-parameter fit in the grid procedure that maximizes the neutral pion
SSA in the large $x_F$ region and applying the error estimate procedure
described in Appendix A of Ref.~\cite{Anselmino:2008sga}. See text for
more details. }
\label{fig:an-star-free7}
\end{figure}

It is not clear how to combine the statistical error band, associated with
the full 7 or 9 free-parameter best fits of SIDIS and $e^+e^-$ data,
with the scan bands. Therefore, in order to understand to what extent
the statistical errors on the best fit parameters may affect the capability
of the Collins effect to reproduce the large $x_F$ STAR data, we have adopted
the following strategy: besides considering the envelope of the full
set of curves produced by the scan procedure, we have considered explicitly
each of these curves, isolating the set leading to the largest asymmetries
in the large $x_F$ region; we have then evaluated, as in Appendix A of
Ref.~\cite{Anselmino:2008sga}, the corresponding statistical error band,
which covers larger values of the asymmetry. Our result is presented in
Fig.~\ref{fig:an-star-free7}. Again, it appears that the Collins effect
alone cannot account for the large $x_F$ data. Notice also that trying to
fit the large $x_F$ data on $A_N$ might lead to an overestimation of the
same data at smaller $x_F$, which have tinier error bars.

\begin{figure}
\begin{minipage}[c]{19cm}
\includegraphics[width=14cm,angle=0]{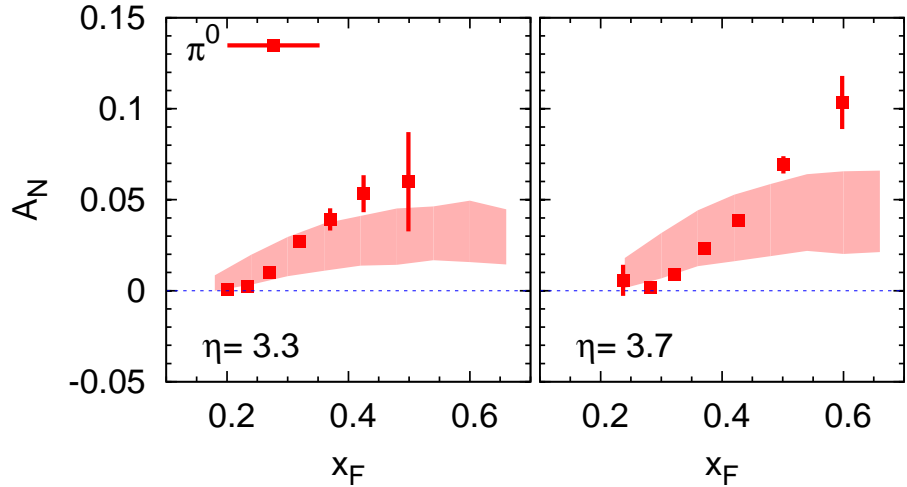}
\end{minipage}
\caption{
The same as for Fig.~\ref{fig:an-star-free7}, but this time with the
statistical error band generated starting from the scan procedure with
11 free parameters. See text for more details.}
\label{fig:an-star-free11}
\end{figure}

There is still another issue that deserves some attention. Although
simplified, our parameterization of the TMDs and of their functional
shape involves in principle a huge number of fit parameters.
Since most of these parameters are highly correlated, adopting larger
set of parameters would lead to larger uncertainties in their value.
Therefore, reasonable fits require a reduction in the number of fit
parameters and involve a careful choice of the most significant ones.
This choice may have consequences on the allowed values of the asymmetries,
particularly for kinematical regions not covered by the data sets
used for the fitting procedure. In the present scan procedure we use a
7-parameter fit and a grid of values for the two additional parameters
$\beta_u$, $\beta_d$. In order to investigate if a larger set of free
parameters for the scan procedure could modify our conclusions about the
Collins effect for the STAR data in the large $x_F$ range, we have
repeated our scan procedure by starting from a preliminary reference fit
with 13 free parameters,
\begin{equation}
N_u^T,\,N_d^T,\,\alpha_u,\,\alpha_d,\,\beta_u,\,\beta_d,\,
N_{\rm fav}^C,\,N_{\rm unf}^C,\,
\gamma_{\rm fav},\,\gamma_{\rm unf},\,\delta_{\rm fav},\,
\delta_{\rm unf},\,M_h\,,
\label{eq:11-par}
\end{equation}
and generating again the scan band on the two dimensional grid for the
fixed $\beta_{u,d}$ parameters by fitting for each grid point the remaining
11 parameters. This naturally results in a sizably larger scan band, with
its upper edge approaching better $A_N$ at the larger $x_F$ values.
However, also in this case, looking at all the 81 fit sets we find that
the curve with the best behaviour at large $x_F$ approaches the upper edge
of the scan band in the full $x_F$ range. It therefore largely misses (overestimates) the lower $x_F$ values of the asymmetry.
In Fig.~\ref{fig:an-star-free11} we present the statistical error band
on the Collins contribution to $A_N(p^\uparrow p \to \pi^0 X)$ generated,
following Appendix A of Ref.~\cite{Anselmino:2008sga}, from the 11-parameter
best-fit set which optimizes the agreement with STAR data at large $x_F$.

Let us finally make some comments on the (charged and neutral) pion SSAs
for the E704 kinematics \cite{Adams:1991rw,Adams:1991cs}. The situation
in this case is complicated by the fact that, contrary to the STAR and
BRAHMS kinematics, the unpolarized cross sections are largely underestimated
within the TMD LO factorized approach when adopting the values of
$\langle k_\perp^2\rangle$ and $\langle p_\perp^2\rangle$ extracted from
SIDIS data. Indeed, much larger effective values are required to reconcile
the TMD estimates with data, as it was shown in Ref.~\cite{D'Alesio:2004up}.
However, this fact should have less influence on the SSAs, defined as ratios
of (sums and differences of) single-polarized cross sections.

We have therefore directly applied the scan procedure, as illustrated
above, also to the E704 results. Again, it turns out that the scan band
could cover the data for the neutral pion SSA, with some problem at the
largest $x_F$ values. However, it largely misses the huge charged pion
SSAs observed in the same kinematical region.

\section{Conclusions}

We have investigated the possible role of the Collins effect in explaining
the large SSAs observed in $\pup p \to \pi X$ reactions; we have done so
within a TMD factorized scheme, and have revisited a previous work on the
same issue, correcting a numerical error and using new experimental data
and new phenomenological information on the transversity distribution and
the Collins function.

We can conclude that, to the best of our present knowledge, based on SIDIS
and $e^+e^-$ data, the Collins effect alone seems to be able to reproduce
the available RHIC data on pion single spin asymmetries in polarized
$pp$ collisions, only in the small Feynman $x$ region, $x_F \lsim 0.3$.
Above that, which is the region where the values of $A_N$ increase, the
Collins effect alone is not sufficient.

Additional mechanisms are required in order to explain the size of the
$A_N$ asymmetry in this region. One can obviously think of the Sivers
effect~\cite{Anselmino:1994tv,Anselmino:1998yz,D'Alesio:2004up,D'Alesio:2007jt}.
Since TMD factorization has not been proven and is still under active debate
for single inclusive particle production in hadronic collisions and since
universality breaking effects are possible, one does not know exactly how
to use the parameterizations of the quark Sivers functions extracted from
SIDIS data in $pp$ collisions. A recent use of the SIDIS Sivers functions
in a collinear higher-twist approach to SSAs in $pp$ collisions -- rather
than in the TMD factorized approach -- has been found to give a sizable
contribution to $A_N$, but with the wrong sign~\cite{Kang:2011hk}.
A similar conclusion holds in a modified generalized parton model approach
with TMD factorization~\cite{Gamberg:2010tj}. Notice that such a problem
does not occur if one simply adopts the SIDIS Sivers functions in the TMD
factorized scheme~\cite{Boglione:2007dm,Anselmino:2009hk}.
Much further investigation is necessary.

\acknowledgments
We are grateful to F. Yuan for helping us in clarifying the sign
mistake in Refs.~\cite{Anselmino:2004ky,Anselmino:2005sh}.
We acknowledge support of the European Community under the FP7
``Capacities - Research Infrastructures'' program (HadronPhysics3,
Grant Agreement 283286). Some of us (M.A., M.B., U.D., F.M.)
acknowledge partial support from MIUR under Cofinanziamento PRIN 2008 and
E.L. from INFN. U.D. is grateful to the Department of Theoretical Physics II
of the Universidad Complutense of Madrid for the kind hospitality extended
to him during the completion of this work.


\end{document}